\documentclass[pra,10pt,twocolumn,superscriptaddress,floatfix]{revtex4-1}
\usepackage{amsmath}
\usepackage{latexsym}
\usepackage{amssymb}
\usepackage{graphics,epstopdf}
\usepackage{hyperref}
\usepackage{color}
\hypersetup{
    colorlinks = true,
    linkcolor =blue,
	citecolor=blue, 
	urlcolor=blue 
}

\newcommand{\ed}{\end{document}}
\newcommand{\beq}{\begin{equation}}
\newcommand{\eeq}{\end{equation}}

\usepackage{mathtools}

\newcommand{\sandwich}[3]{\langle #1 \vert #2 \vert #3 \rangle}

\newcommand{\Tr}{\ensuremath{\mathrm{Tr}\,}}
\newcommand{\avg}[1]{\ensuremath{\langle #1 \rangle}}
\newcommand{\ket}[1]{\ensuremath{\vert #1 \rangle}}
\newcommand{\bra}[1]{\ensuremath{\langle #1 \vert}}

\MakeRobust{\eqref}
\newcommand{\eqnref}[1]{Eq.~\eqref{#1}}
\newcommand{\figref}[1]{Fig.~\eqref{#1}}
\newcommand{\hs}{\hat{\sigma}}
\newcommand{\hc}{\hat{c}}
\newcommand{\hcd}{\hat{c}^{\dagger}}
\newcommand{\heta}{\hat{\eta}}
\newcommand{\hetad}{\hat{\eta}^{\dagger}}
\newcommand{\hrho}{\hat{\rho}}
\newcommand{\Hop}{\hat{H}}
\newcommand{\Vop}{\hat{V}}
\newcommand{\Dop}{\hat{D}}
\newcommand{\Aop}{\hat{A}}

\newcommand{\hsb}{{\hat{\pmb{\sigma}}}}
\newcommand{\esb}{\pmb{\sigma}}
\newcommand{\sz}{\sigma_z}

\begin{document}
\title{Read-out of Quasi-periodic Systems using Qubits}
\author{Madhumita Saha}
\email{madhumita.s@iitgn.ac.in}
\affiliation{Indian Institute of Technology Gandhinagar, Palaj, Gujarat 382355, India}
\author{Bijay Kumar Agarwalla}
\email{bijay@iiserpune.ac.in}
\affiliation{Department of Physics, Indian Institute of Science Education and Research Pune, Dr. Homi Bhabha Road, Ward No. 8, NCL Colony, Pashan, Pune, Maharashtra 411008, India}
\author{B. Prasanna Venkatesh}
\email{prasanna.b@iitgn.ac.in}
\affiliation{Indian 
Institute of Technology Gandhinagar, Palaj, Gujarat 382355, India}

\begin{abstract}
We develop a theoretical scheme to perform a read-out of the properties of a quasi-periodic system by coupling it to one or two qubits. We show that the decoherence dynamics of a single qubit coupled via a pure dephasing type term to a 1D quasi-periodic system with a potential given by the Andr{\'e}-Aubry-Harper (AAH) model and its generalized versions (GAAH model) is sensitive to the nature of the single particle eigenstates (SPEs). More specifically, we can use the non-markovianity of the qubit dynamics as quantified by the backflow of information to clearly distinguish the localized, delocalized, and mixed regimes with a mobility edge of the AAH and GAAH model and evidence the transition between them. By attaching two qubits at distinct sites of the system, we demonstrate that the transport property of the quasi-periodic system is encoded in the scaling of the threshold time to develop correlations between the qubits with the distance between the qubits. This scaling can also be used to distinguish and infer different regimes of transport such as ballistic, diffusive and no transport engendered by SPEs that are delocalized, critical and localized respectively. In addition, the localization length of the SPEs can also be gleaned from the exponential decay of correlations at long times as a function of distance between qubits. When there is a mobility edge allowing the coexistence of different kinds of SPEs in the spectrum, such as the coexistence of localized and delocalized states in the GAAH models, we find that the transport behaviour and the scaling of the threshold time with qubit separation is governed by the fastest spreading states.
\end{abstract}

\maketitle
\section{Introduction}
Quantum simulation has at its heart the idea of using simple, controllable systems to study other complex, less controllable quantum systems. This idea has now matured into an active area of current research \cite{Georgescu2014,Quantumsimulation1,Quantumsimulation2} spread over multiple quantum technology platforms such as ultracold atoms \cite{Jaksch2005,Bloch2008}, superconducting qubits \cite{Houck2012}, trapped ions \cite{Blatt2012}, and exciton-polariton condensates in microcavities \cite{Hartmann2006,Greentree2006,Rodriguez2016}. Along with progress towards the ultimate goal of simulating strongly correlated many-body systems, experimental realisations of quantum simulation, especially in ultracold atoms, have provided the first clean and clear realizations of textbook solid-state phenomena such as Bloch oscillations \cite{BenDahan1996} and Anderson localization \cite{Billy2008,Roati2008}. These demonstrations have not only shed light on the phenomena but the degree of control and manipulation possible with the experimental set-ups has further stimulated research in related directions \cite{Modugno2010,Eckardt2017}. Furthermore, the high degree of control in such quantum technology platforms has also led to novel ways to probe and measure the system. In addition to standard projective measurements realized for example in the absorption imaging of ultracold atomic gases \cite{Bloch2008} or in the famous quantum gas microscope experiments \cite{Bakr2009}, there is now a large body of work on non-destructive, in-situ measurements such as using the light-field of a cavity coupled to atoms to probe them \cite{Northup2005,Ritsch2013} and using impurities immersed in ultracold atomic systems as probes \cite{Klein2005,Klein2007,Haikka2013,Ratschbacher2013,Scelle2013,Haikka2014,Cetina2016,Kleinbach2018,Schmidt2018,Schmidt2019,Hangleiter2015,Elliott2016,Streif2016,Usui2018,Borrelli2013}. The common feature in such measurement techniques is the coupling of the system of interest to an ancillary probe system and performing usual projective measurements on the probe. In many situations the tailoring of system-probe interactions and/or continuous measurement of the probe leads to interesting advantages in terms of the observables that can be accessed \cite{Mekhov2007} and measurement precision \cite{Peden2009, PrasannaVenkatesh2009,  Kebler2016, Hosten2016, Hosten2016a}. The second attribute of these approaches is that such measurements are essentially a controlled way to `open' up an otherwise `closed' quantum system leading to richer behaviour \cite{Ritsch2013} and new avenues of quantum control \cite{Mekhov2009}. In line with this, in the present article we develop a theoretical scheme to read-out the properties of a non-interacting quasi-periodic system by coupling it to two-level `probe' qubit systems.

Quasi-periodic lattice systems are in a sense in-between ordered systems with periodic on-site potential and completely disordered systems with random on-site potential. The latter provide the setting for the famous Anderson localization phenomenon with localized single particle states in 1D and 2D, and a possible coexistence of localized and delocalized states with a mobility edge in 3D. In contrast, quasi-periodic systems have rich localization and transport properties even in 1D. An exemplary system to illustrate this is the one with the Andr{\'e}-Aubry-Harper (AAH) potential, where by tuning the system parameters it is possible to get delocalized, critical and localized single particle states \citep{Aubry}. Moreover, various generalized versions of the AAH potentials that support a single-particle mobility edge even in 1D have been proposed \citep{Commensurate_AAH,GAAH_mobility_edge}. The intense theoretical research in this area focusing on different properties such as single-particle and many body localization, transport etc. has been complemented and invigorated by the experimental realizations of such quasi-periodic potentials and the detection of localization and other associated phenomena \citep{I_bloch_experiment,expt1,expt2,expt3,zilberberg1,zilberberg2,Goblot2020,GAAH_experiment,Nagler2020}. While typical theoretical works have largely focused on localization and transport properties of the isolated quasi-periodic systems \citep{PhysRevLett.115.230401,PhysRevB.99.224204, Laurent}, in some recent work these systems have also been treated in an `open' quantum manner by attaching two baths at the edges of the lattice and studying the non-equilibrium steady state transport properties of the system \citep{PhysRevB.99.224204,Archak_AAH,Archak_phase_diagram,Archak_AAH1,PhysRevE.96.032130,PhysRevB.99.035143}. Our aim is to take a middle path where we will couple the quasi-periodic system to qubits, thereby in a controlled manner `opening' the system with the goal of inferring the localization and transport properties of the system from the dynamics of the qubits. 

Since our system of interest, to which the qubits are to be coupled, is a simple non-interacting quadratic model, it can be easily diagonalized (numerically). Hence, from the point of view of the qubits the system is essentially a bosonic bath with energy of the modes given by the single particle eigenenergies of the quasi-periodic system. In this light, we will use known results concerning spin-Boson type models especially the so called dephasing bath models as we will consider a form of coupling between the qubits and quasi-periodic system that commutes with the free hamiltonian of the qubits \cite{MassimoPalma1996,Lidar2001,Reina2002,Breuer2007,Cipolla2018,Chen2018,Milazzo2019}. Thus, when a single qubit is coupled to the system it results in the pure dephasing or decoherence of the qubit state. We will show that the decoherence dynamics of this qubit is greatly influenced by the nature of single particle states of the quasi-periodic system and can hence be used as a read-out of the same. More specifically, we will see that the degree of non-markovianity of the qubit dynamics will provide a clear quantitative measure for the read-out that correlates with the localization properties of the system. Subsequently, we will show that we can access very different aspect of the quasi-periodic system's properties, namely transport properties, by studying the dynamics of the correlation between two qubits coupled to different sites of the system. 

Before discussing the structure of the article, we would like to briefly draw attention to some relevant previous work. We note that the idea of using qubits as probes of quantum systems is a topic that has received significant attention especially in the context of ultracold atoms with contributions both on the theoretical \cite{Klein2005,Klein2007,Goold2011,Knap2012,Haikka2013,Borrelli2013,Haikka2014,
Hangleiter2015,Elliott2016,Streif2016,Usui2018,Francesco1,Francesco2,Francesco3,Francesco4,Francesco5} and experimental \cite{Ratschbacher2013,Scelle2013,Schmidt2018,Kleinbach2018,Cetina2015,Cetina2016} fronts. Of particular relevance to the present article are theoretical works considering the decoherence dynamics of qubits coupled to ultracold Fermi gases \cite{Goold2011,Cetina2015} as well as performing thermometry using such qubits \cite{Mitchison2020}, using a qubit probe to study phononic excitations of a ultracold Bose gas \cite{Hangleiter2015}, coupling of multiple qubits to an ultracold atoms in a lattice to study local fluctuations and two-point correlations of density \cite{Elliott2016,Streif2016}, using single qubit to read-out the local excitation spectrum of a quantum gas in a two dimensional optical lattice \cite{Usui2018}. These theoretical developments have been complemented by a range of related experiments such as the study of the decoherence of a single ionic qubit immersed in a cold Bosonic gas \cite{Ratschbacher2013}, decoherence of a Fermionic atom in a double well potential (forming a 'motional' qubit) immersed in a Bosonic gas \cite{Scelle2013}, interferometric measurements of a qubit to probe the many body dynamics of an ultracold Fermi gas \cite{Cetina2016}. In addition to such system specific studies of qubits as probes, there have also been theoretical proposals for reaching ultimate limits of thermometry using dephasing dynamics of a qubit \cite{PhysRevA.91.012331,PhysRevLett.114.220405} and experimental implementation of such thermometry using photon-polarization qubits \cite{Tham2016}. Another important feature of our article that has received attention previously is the idea to use  non-markovianity of the qubit dynamics as a read-out mechanism \cite{Haikka2013,Haikka2014,Lorenzo2017,Giorgi2020}. In particular the proposal to evidence a structural phase transition in a trapped-ion chain using the non-markovian dynamics of one of the ionic qubits \cite{Borrelli2013} in the chain and the study \cite{Giorgi2019} suggesting to probe a topological phase transitions in the Su-Schreiffer-Heeger model, by measuring the non-markovianity in the decoherence dynamics of a coupled qubit are similar in spirit to our results to use non-markovianity to probe the localization-delocalization transition in the bosonic AAH model. In the same vein, in \cite{Francesco1} the coupling of a qubit to a Fermionic quasi-periodic lattice system was considered and it was shown that the localized phase can be considered as a non-markovian environment for the qubit. The form of the coupling we will use and the focus on reading out transport properties using multiple qubits coupled to bosonic AAH and generalized AAH lattice systems distinguishes our work from \cite{Francesco1}.

The paper is organized as follows. In Sec. II, we describe the model and the basic theoretical framework for the problem starting from a recapitulation of the essential properties of the quasiperiodic AAH and generalized AAH models, followed by the read-out scenarios with one and two qubit coupled to the quasiperiodic system. In Sec. III we will describe the main results that will establish how the dynamics of the coupled qubits can be used as a read out of the properties of the system. In Sec. IV we conclude the paper and comment on the possible experimental realization of the ideas we have presented. We present the details regarding the calculation of decoherence rates of the qubit in the Appendix.

\section{Model and theory}
\begin{figure}
	\includegraphics[width=\linewidth]{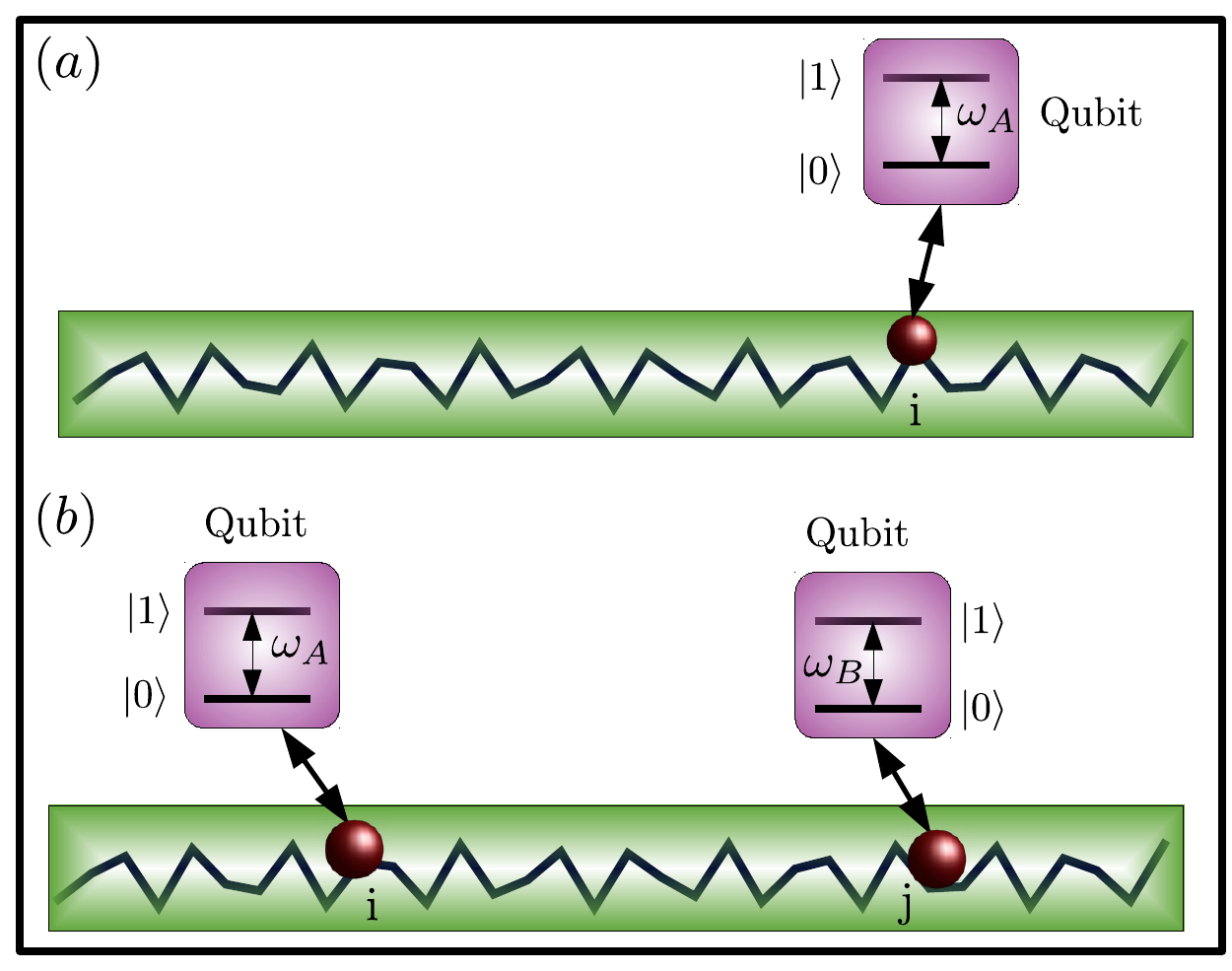}
	\caption{(Color online). Schematic representation of the read-out of a quasi-periodic system using one or two qubits. (a) A single qubit (represented by the red circle) with energy splitting $\omega_A$ between its two levels is coupled to the $i$ th site of an one-dimensional quasi-periodic system (the black lines represent the on-site potential of the lattice). (b) Two qubits with splitting energies $\omega_A$ and $\omega_B$ are coupled to two different sites $i$ and $j$ of the quasi-periodic system respectively.} 
	\label{fig:figure1}
\end{figure}
The central claim of our paper is that the nature of on-site potential, localization and transport properties of an otherwise isolated non-interacting system can be read out by coupling the system to one or two-qubit probes. We will demonstrate this claim by considering a specific 1-d tight-binding model - the Andr{\'e}-Aubry-Harper (AAH) and its generalized versions. In Fig. \ref{fig:figure1}, we depict in a schematic manner the coupling scheme we will use for the read-out. We consider two scenarios, in the first case a single qubit [\figref{fig:figure1} (a)] will be coupled to a given site of the 1-d tight-binding system and in the the second case two qubits will be coupled at two different sites [\figref{fig:figure1} (b)]. We will see that the two scenarios lead to the read out of complementary aspects of the 1-d tight-binding system. Towards this, we begin by briefly revisiting the basic properties of AAH and the generalized Andr{\'e}-Aubry-Harper (GAAH) models relevant to our study, followed by the details of the one and two-qubit coupling model that we will employ and the read-out mechanism. 

\subsection{Properties of 1D AAH and GAAH Models}
A paradigmatic example of quasi-periodic system that has been the subject to intense theoretical \cite{Pandit83,Commensurate_AAH,Archak_AAH,Archak_phase_diagram,GAAH_mobility_edge,PhysRevLett.115.230401,PhysRevB.99.224204} and experimental research \cite{I_bloch_experiment,expt1,expt2,expt3,zilberberg1,zilberberg2,GAAH_experiment} is the 1D tight-binding nearest neighbour hopping model with Andr{\'e}-Aubry-Harper(AAH) potential and its generalized version. The Hamiltonian for the 1D $N$-site tight-binding chain with GAAH potential has the form \citep{GAAH_mobility_edge} (we take $\hbar = 1$ throughout)
\begin{align}
\hat{H} & = \sum_{n=1}^N \left( \mu+\frac{2 \lambda \cos[2\pi b n +\phi]}{1+\alpha \cos[2\pi b n +\phi]} \right ) \hcd_n \hc_n \nonumber \\ 
& + \sum_{n=1}^{N-1} \left( \hcd_n c_{n+1} + \hcd_{n+1} \hc_n \right) \, .
\label{eq:GAAH}
\end{align}
{Here we have written the hamiltonian in units of the hopping strength $J$, $\hcd_n (\hc_n)$ is the bosonic creation (annihilation) operator at site $n$, $\mu$ is a constant on-site potential energy, $b$ is an irrational number which makes the potential quasi-periodic, and $\phi$ is the phase factor which also determines the different configurations of the quasi-periodic potential. Moreover henceforth we take the phase $\phi=0$ and choose $\alpha,\lambda>0$ for simplicity. For $\alpha=0$, the GAAH hamiltonian reduces to the regular AAH model. When $\alpha=0$, for any choice of irrational number $b$ all the single particle eigenstates (SPE) are completely delocalized for $\lambda<1$, exponentially localized for $\lambda>1$ \cite{Aubry}. For $\lambda = 1$ all the SPEs are critical. Thus for the AAH model, by changing $\lambda$ one can access different kinds of SPE. In contrast, the GAAH model with $\alpha, \lambda >0$ supports a mobility edge at the energy $ E=\mu + 2 (1- \lambda)/\alpha$ \cite{GAAH_mobility_edge}. If the energy $E$ falls within the spectrum, all SPEs with energy less than $E$ are extended while those higher than $E$ are localized. We would like to reiterate that henceforth all parameters with the dimensions of energy such as $\mu,\lambda$ (including the qubit parameters to be introduced in the following sub-section) are understood to be in units of $J$ and variables with the dimensions of time will be in units of $J^{-1}$.

\subsection{One qubit coupled to a quasi-periodic system}
Our first strategy to read out the properties of the GAAH chain \eqnref{eq:GAAH} is to couple a qubit of frequency $\omega_A$ with the site $i$ in the chain resulting in a total hamiltonian of the form:
\begin{align}
\label{eq:totalhamil_onequb}
\hat{H}_{1q}=\hat{H} + \frac{\omega_A}{2} \hs_z^i + g \hs_z^i (\hc_i^{\dagger}+\hc_i),
\end{align}
with $g$ denoting the strength of the coupling between the qubit and the chain. We can rewrite the above hamiltonian in terms of the eigenmodes of the GAAH chain \emph{i.e.} in the representation that diagonalizes the hamiltonian \eqnref{eq:GAAH} as $\Hop = \sum_{k=1}^{N} \omega_k \hetad_k \heta_k$, with $\omega_k$ denoting the $k^{\mathrm{th}}$ single particle eigenenergy and $\heta_k$ the corresponding annihilation operator. The unitary matrix that transforms between the local operators at site $i$ and the eigenmodes, denoted by $S$, satisfies:
\begin{align}
\hc_i = \sum_{k=1}^{N} S_{i,k} \heta_k. \label{eq:transform}
\end{align}
With this transformation, we can recast \eqnref{eq:totalhamil_onequb} as:
\begin{align}
\label{eq:spinboson_1q}
\hat{H}_{1q}^{SB}=\frac{\omega_A}{2} \hs_{z}^i + \sum_{k=1}^{N} \omega_k \heta_k^{\dagger} \heta_k + \sum_{k=1}^{N}  \hs_{z}^i (g_k^i\heta_k^{\dagger}+ g_k^{i*} \heta_k),
\end{align}
with $g_k^i=g S_{i,k}$, and $\hs_{\tau}^{i}$ with $\tau = {z,\pm}$ denoting the usual Pauli and ladder operators for the qubit. From the above equation it is clear that the hamiltonian of interest here falls into the familiar family of spin-Boson type hamiltonians and more specifically the spin-Boson problem with a dephasing bath that has been extensively studied especially in the context of decoherence in quantum computation \cite{MassimoPalma1996,Lidar2001,Reina2002,Breuer2007,Cipolla2018,Chen2018,Milazzo2019}.The key advantage of this model is its exact solubility which arises from the fact that the qubit hamiltonian $\frac{\omega_A}{2} \hs_{z}^i$ commutes with the term representing the interaction with the bosonic modes. For the sake of completeness, we provide the details of this calculation (for both the one and two qubit situations) in the Appendix \ref{app:dephSBmodel} and present only the results here.

Since we aim to use the qubit to read out the properties of the GAAH chain, the key quantity of interest is the state of the qubit at a given time $t$. Working in the Heisenberg picture, we first note that due to the nature of the coupling between the qubit and the Boson the population of the qubit states is unaffected under the dynamics generated by \eqnref{eq:spinboson_1q} \emph{i.e.} $\hs_z(t) = \hs_z(0)$. Thus, the qubit dynamics is completely described by the time dependence of the spin-ladder or coherence operators $\hs_{\pm}(t)$ and is given by:
\begin{align}
\hs_{-}^{i}(t) = e^{-i\omega_A t} \hs_{-}^i(0) \otimes \prod_{k=1}^N \Dop_{k}(\alpha_k^i), \label{eq:singlequbitSigma}
\end{align}
with $\alpha_k^i = \frac{2g_k^i(1-e^{i\omega_k t})}{\omega_k}$ and $\Dop_{k}(\alpha) = \exp(\alpha \hetad_k(0) -\alpha^{*} \heta_k(0)$ is the displacement operator for the $k^{\mathrm{th}}$ bosonic mode. We choose the initial time density operator for the spin-Boson system as the product state $\hrho(0) = \hrho_{s}(0) \otimes (e^{-\beta \Hop}/Z_{\beta})$. Here $Z_{\beta} = \Tr_{B}[e^{-\beta \Hop}]$ denotes the partition function corresponding to a thermal state of the GAAH chain with inverse temperature $\beta = 1/T$ (taking $k_B = 1$) and the trace is with respect to the states of the GAAH chain. In addition, we choose the initial state of the qubit as a pure state \emph{i.e.} $\hrho_s(0) = \ket{\psi_0}^{i}{}^{i}\bra{\psi_0}$ with non-zero initial coherence $\avg{\hs_{-}^i(0)}$. With this setting we obtain,
\begin{align}
\avg{\hs_{-}^{i}(t)} & = \avg{\hs_{-}^i(0)} e^{-i\omega_A t -\Gamma_i (t)}, \label{eq:cohevol1qubit} \\
\Gamma_i(t)&=4\sum_{k=1}^{N} |g_k^i|^2 \coth\left(\frac{\beta \omega_k}{2}\right)\frac{1-\cos(\omega_k t)}{\omega_k^2}.\label{eq:dc}
\end{align}
Thus the interaction with site $i$ of the bosonic GAAH chain leads to a decay of the qubit coherence at the rate $\Gamma_i(t)$-which we will henceforth refer to as the decoherence factor. In order to distinguish the contribution of the thermal and quantum excitations of the bosonic chain to the decoherence factor we first note that
$\coth\left(\frac{\beta \omega_k}{2}\right) = 1 + 2 \bar{n}(\beta,\omega_k)$ where $\bar{n}(\beta,\omega_k)=(e^{\beta \omega_k} -1)^{-1}$ is the average thermal occupation of the $k^{\mathrm{th}}$ mode. This allows us to write $\Gamma_i(t)=\Gamma_{i,\mathrm{vac}}(t)+\Gamma_{i,\mathrm{th}} (t)$, with 
$\Gamma_{i,\mathrm{vac}}(t) = \sum_{k=1}^{N} \frac{\vert \alpha_k^i \vert^2}{2}$ and 
$\Gamma_{i,\mathrm{th}}(t) = \sum_{k=1}^{N} \bar{n}(\beta,\omega_k) \vert \alpha_k^i \vert^2$. 

As it is apparent from \eqnref{eq:dc} that the decoherence factor $\Gamma_i(t)$ encodes information about the GAAH chain via the magnitude of the coefficients $|g_k^i|^2$, the first mechanism we would like to propose for the read-out is to measure this decoherence factor as a function of time and the coupling site $i$. We will show in detail in Sec. \ref{sec:Results} that while the site dependence of the decoherence factor at a given time will reflect the site-to-site variation of the applied potential, the dynamics, irrespective of which site we couple to, will provide a clear indication of the single particle localization properties of the chain.

\subsection{Two qubits coupled to a quasi-periodic system}

A standard way to understand the transport properties of an isolated system is to study the diffusion dynamics of an initial wave packet 
\citep{diffdynamics,PhysRevA.80.053606,PhysRevA.87.023625,Archak_AAH} 
Starting with a spatially localized initial wave packet, the scaling of the wave packet's width $w$ as a function of evolution time $t$ provides a clear indicator of the transport properties to be expected. Assuming a generic scaling of the form $w \sim t^{\eta}$, we have that for ballistic transport $\eta=1$, for localized case $\eta=0$ and for diffusive transport $\eta=0.5$. $0<\eta<0.5$ leads to sub-diffusion and $0.5<\eta<1$ leads to super-diffusion. Indeed, in the first experimental realization of the AAH model with ultracold atomic systems \cite{Roati2008}, the behavior of the wavepacket spreading was precisely used to evidence localization. Inspired by this, in our second scheme for the read out of the GAAH chain we consider [see \figref{fig:figure1} (b)] two qubits of frequency $\omega_A$ and $\omega_B$ coupled simultaneously to two different sites $i$ and $j$ respectively of the quasi-periodic GAAH chain. We will show that the dynamics of correlations between the qubits coupled to different sites of the chain can be used to extract the scaling coefficient $\eta$ characterizing the nature of transport.

The hamiltonian for the GAAH chain with two qubits attached can be written in a manner similar to \eqnref{eq:spinboson_1q} as:
\begin{align}
\hat{H}_{2q}^{SB}&=\frac{\omega_A}{2} \hs_{z}^i + \frac{\omega_B}{2} \hs_{z}^j + \sum_{k=1}^{N} \omega_k \heta_k^{\dagger} \heta_k  \nonumber \\
&+ \sum_{k=1}^{N}  \hs_{z}^i (g_k^i\heta_k^{\dagger}+ g_k^{i*} \heta_k) + \sum_{k=1}^{N}  \hs_{z}^j (g_k^j\heta_k^{\dagger}+ g_k^{j*} \heta_k) \label{eq:spinboson_2q}
\end{align}
which is still in the class of spin-boson models with dephasing baths. As a result, the dynamics of this model can again be exactly solved \cite{Lidar2001,Cipolla2018,Chen2018} and we present the details in the Appendix \ref{app:dephSBmodel}. As before, working in the Heisenberg picture, we note that due to the nature of the coupling we have that $\hs_z^i(t) = \hs_z^i(0), \hs_z^j(t) = \hs_z^j(0)$, and the only non-trivial dynamics is for the ladder operators of the individual qubits and is given by:
\begin{align}
\hs_{\pm}^{i}(t) = [e^{\pm i\omega_A t} \hs_{\pm}^i(0)] \otimes e^{ \pm i \Delta \Omega_{-} (t) \hs_z^j} \otimes  \prod_{k=1}^N \Dop_{k}(\pm \alpha_k^i) \label{eq:ladderifnt},\\
\hs_{\pm}^{j}(t) = e^{ \pm i \Delta \Omega_{+} (t) \hs_z^i} \otimes [e^{\pm i\omega_B t} \hs_{\pm}^j(0)] \otimes  \prod_{k=1}^N \Dop_{k}(\pm \alpha_k^j) \label{eq:ladderjfnt},
\end{align}
with the Lamb-shift term given by 
\begin{align}
\Delta \Omega_{\pm}(t) = \sum_{k=1}^N \frac{4}{\omega_k^2} \left( \left[\sin(\omega_k t)-\omega_k t \right]\operatorname{Re}[g_k^i g_k^{j*}] \nonumber \right. \\  \left . \pm \left[ 1- \cos(\omega_kt)\right]\operatorname{Im}[g_k^i g_k^{j*}]\right) \label{eq:twoqLambshift}.
\end{align}
As mentioned above, our interest is in calculating the equal time correlation between the qubit observables coupled to the GAAH chain at different sites. As a particular measure of the correlation, we would like to examine the dynamics of the covariance between the operators $\hs_{-}^i(t)$ and $\hs_{+}^j(t)$ defined as $\operatorname{Cov}(\hs_-^i \hs_+^j) \equiv \langle \hs_-^i(t) \hs_+^j (t) \rangle - \langle \hs_-^i(t) \rangle \langle \hs_+^j (t) \rangle$. Here the averages are taken with respect to the initial state of the GAAH chain and qubits which, as before, we take as the product state $\hrho(0) = \hrho_{s}^{i}(0) \otimes \hrho_{s}^{j}(0) \otimes (e^{-\beta \Hop}/Z_{\beta})$. Here, $\hrho_{s}^{i}(0)$ [$\hrho_{s}^{j}(0)$] corresponds to the initial state of the qubit coupled to the $i^{\mathrm{th}}$ [$j^{\mathrm{th}}$] site. Using Eqs. \eqref{eq:ladderifnt} and \eqref{eq:ladderjfnt} we can write the covariance as:
\begin{align}
&\operatorname{Cov}(\hs_-^i \hs_+^j) = e^{-i(\omega_A-\omega_B)t} \left (\avg{\hs_-^i(0) \hs_+^j(0)} e^{-\Gamma_{ij}(t)} \right . \nonumber\\ &\left . - \avg{\hs_-^i(0)} \avg{\hs_+^j(0)} \avg{e^{-i\Delta \Omega_{-}(t)\hs_z^j+i\Delta \Omega_{+}(t)\hs_z^i}} e^{-\left[\Gamma_i(t)+\Gamma_j(t)\right]}\right), \label{eq:CovExpression}
\end{align}
with the individual decoherence factors for the two qubits $\Gamma_i(t),\Gamma_j(t)$ given by \eqnref{eq:dc} and the correlated decoherence factor $\Gamma_{ij}(t)$ reads:
\begin{align}
\Gamma_{ij} (t)= 4\sum_{k=1}^{N} |g_k^i-g_k^j|^2 \coth\left(\frac{\beta \omega_k}{2}\right)\frac{1-\cos(\omega_k t)}{\omega_k^2} \label{eq:corrdc}
\end{align} 
From the expression for the covariance \eqnref{eq:CovExpression}, considering a situation with no initial correlation \emph{i.e.}  $\avg{\hs_-^i(0) \hs_+^j(0)} - \avg{\hs_-^i(0)}\avg{\hs_+^j(0)} = 0$, we see that the dynamics generated by the coupling to the GAAH chain will generate correlations. Moreover, since the GAAH chain has only nearest neighbour tunneling, we can anticipate that farther the separation of the two sites to which the qubits are coupled the longer it takes for mutual correlations to develop. Indeed we will see that there is a threshold time-scale $t^*$, determined by the separation between the qubits, before which the correlation is exponentially small. Uptil this threshold time $t^{*}$, the two qubits evolve independently. Moreover, we will see, from the results presented in detail in Sec. \ref{sec:Results}, that the scaling behaviour of $t^{*}$ as a function of the distance between the qubits $\vert i - j \vert$ will serve as a read-out of the transport behaviour expected in the GAAH chain. More specifically we will see that for ballistic transport $t^*$ scales as $|i-j|$, for localized case $t^*$ goes as $exp(|i-j|)$, and  for diffusive transport $t^*$ scales as $|i-j|^2$. In this sense we can think of the distance between the qubits $\vert i - j \vert$ as analogous to the width of wavepacket in the transport analysis using diffusion dynamics discussed at the beginning of this sub-section. While we have chosen here to describe the covariance of the operators $\hs_-^i,\hs_+^j$ to read out the transport properties, we have also verified that the behaviour we describe above is generic for other operator correlations as well. Finally, we remark that for the pathological case when the initial qubit states $\hrho_{s}^{i,j}(0)$ have no coherence \emph{i.e.} $\avg{\hs_-^i(0)}=\avg{\hs_+^j(0)}=0$, then the covariance remains zero for all time and the readout we propose will not work. 


\section{Results} \label{sec:Results}
Having expounded the theoretical underpinnings of our proposed method to read out the properties of the GAAH chain by coupling to qubit(s), we will now present the results that demonstrate our idea. We will divide our results, following the previous section, into two parts. In the first part, we will consider the situation of a single qubit coupled to a given site and show how the properties of the single particle eigen states of the quasi-periodic system can be obtained from the decoherence dynamics of the qubit. In the second part, we will present results illustrating the situation with two qubits coupled at two different sites and extracting a read out of the the isolated system transport properties from the dynamics of their correlation. The results we present are independent of the qubit frequency. While this is obvious from \eqnref{eq:dc} for the single qubit case, in the two qubit case our choice of $\omega_A = \omega_B$ ensures the same for \eqnref{eq:CovExpression}. In our calculations, we will always take $\mu=6$ and $\beta = 1$ unless specified otherwise. The initial state of the qubit(s) is chosen as a pure state $\ket{\psi_0}^{\tau} = (\ket{-1}^{\tau}+\sqrt{2}\ket{1}^{\tau})/\sqrt{3}$ with $\hs_z^\tau \ket{\pm 1}^{\tau} = \pm \ket{\pm 1}^{\tau}$ \emph{i.e.} $\hrho_{spin}^\tau (0)=\ket{\psi_0}^{\tau}{}^{\tau}\bra{\psi_0}$
with $\tau = i,j$. Finally, we note from Eqs. \eqref{eq:dc} and \eqref{eq:corrdc} that the parameter $g$, that quantifies the coupling strength between the qubits and the GAAH chain, enters only as a scaling factor via the coefficients $g_k^i = g S_{i,k}$. As a result, we found that the value of $g$ does not qualitatively affect the results we present and we set $g=1$, in what follows, for simplicity. While we chose $b=\frac{\sqrt{5}-1}{2}$ (the inverse of Golden mean number, where Goleden mean =$\frac{\sqrt{5}+1}{2}$) in our calculations throughout, we have verified that our results will be qualitatively the same for any irrational number $b$.

\begin{figure}
\includegraphics[width=\linewidth]{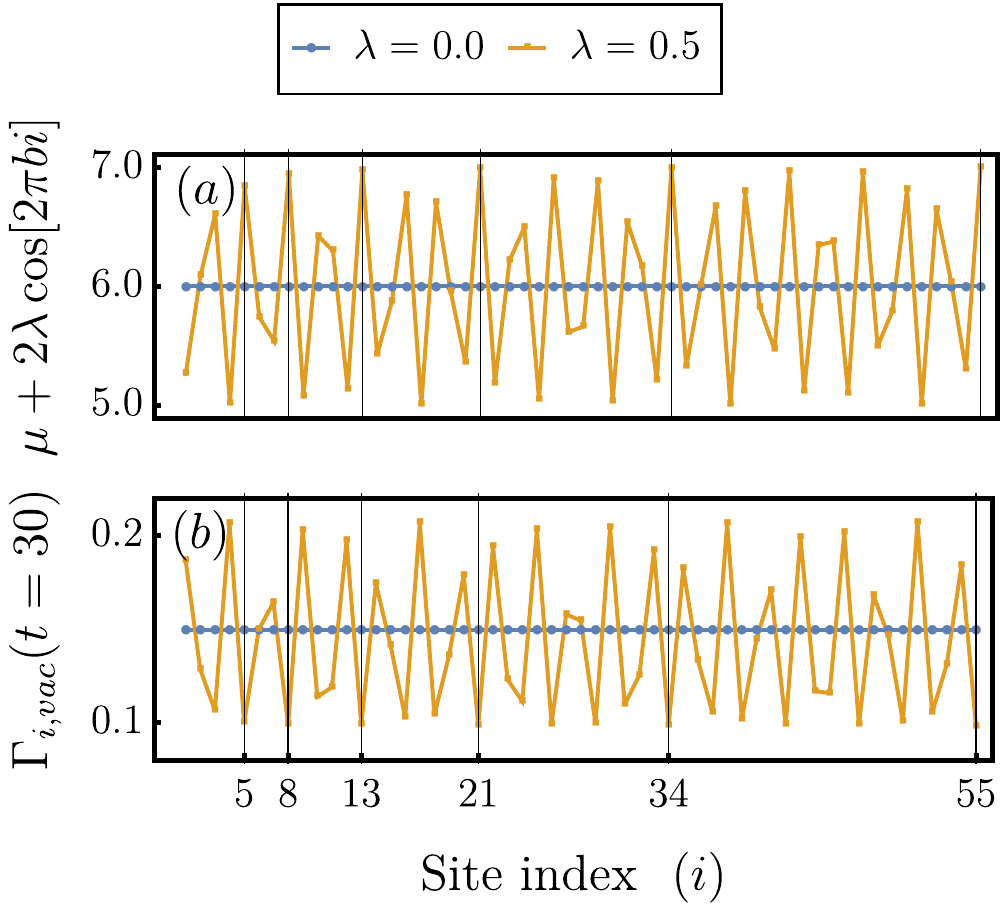}
\caption{(Color online). (a) Ordered ($\lambda = 0$, blue) and quasi-periodic AAH on-site potential ($\lambda = 0.5$, orange) (see \eqnref{eq:GAAH}) as a function of site index  (b). Vacuum decoherence factor for a single qubit as a function of the site to which it is coupled reflecting the underlying on-site potential (total number of sites = $55$). Vertical lines denote the Fibbonacci numbers at which point the potential and decoherence factors almost repeat themselves, as expected for the irrational value $b = \frac{\sqrt{5}-1}{2}$.}
\label{fig:figure2}
\end{figure}
\begin{figure}
	\includegraphics[width=\linewidth]{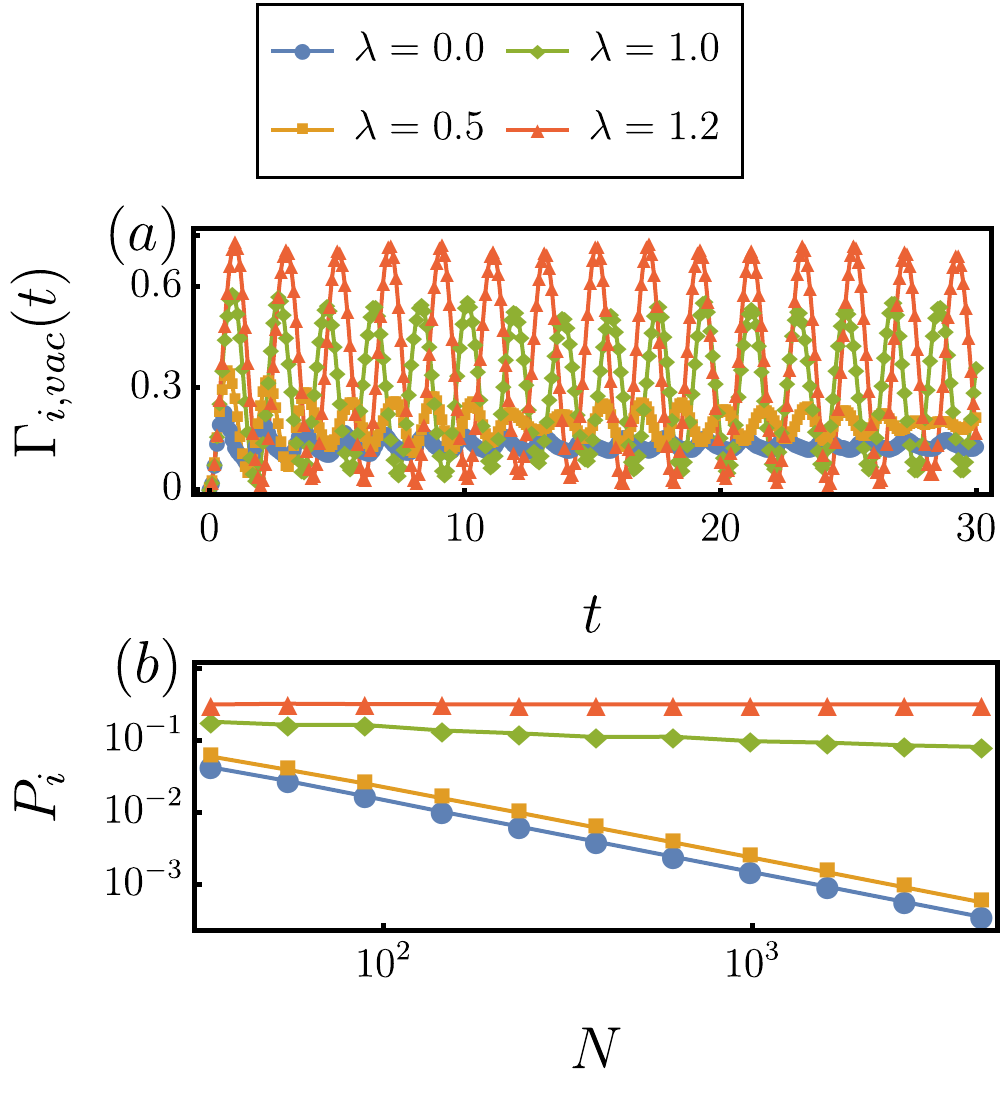}
	\caption{(Color online). (a). Vacuum decoherence factor as a function of time for a $610$-site regular AAH model ($\alpha = 0$) with the qubit coupled to site $i=72$. The dynamics goes from damped to oscillatory as $\lambda$ is tuned over delocalized ($\lambda=0$ and $\lambda=0.5$), critical ($\lambda$=1.0) and localized ($\lambda=1.2$) regimes of the AAH model. (b). Inverse participation ratio of eigenmodes at the qubit coupling site $i$, $P_i$ (see text for definition) as a function of the number of sites $N$. $P_i$ scales as $1/N$ in the delocalized regime, $N^0$ for the localized case, and as $N^{-b}$ with $0<b<1$ in the critical regime.} 
	\label{fig:figure3}
\end{figure}

\subsection{One qubit coupled to a quasi-periodic system}

The first result we present concerns the behaviour of the decoherence function $\Gamma_i(t)$ of a single qubit coupled to the GAAH chain as a function of the choice of the coupling site $i$ at a fixed time $t$. As we can see from \eqnref{eq:dc}, $\Gamma(t)$'s site dependence arises from the term $|g_k^i|^2$ and remains even after the sum over different eigenstates $k$. In this sense, we expect that the decoherence factor at a given time will reflect the nature of on-site potential. For instance, in the tight-binding model with periodic boundary condition, $\vert g_k^i \vert^2=\frac{1}{N}$ is site independent. As a result the decoherence factor will also be site-independent at all times. On the other hand, as we show in \figref{fig:figure2} for the AAH model, the decoherence factor clearly reveals the underlying quasi-periodicity of the potential. There we plot only the vacuum part (zero temperature) of decoherence factor but we find that the behaviour is insensitive to temperature.  Thus, by coupling the qubit to different sites, one can extract the features of the quasi-periodicity in the chain. Since in controlled experimental settings \cite{Roati2008,expt1,expt3} the on-site potential is exactly known this is not necessarily a very useful read-out but nonetheless it serves as a sanity check and sets the stage for the central results we present next.

Having described how the nature of the quasi-periodic potential can be gleaned from the decoherence factor by coupling to different sites of the lattice, our next set of results will illustrate how the nature and properties of the single-particle eigenstates (SPEs) is encoded into the dynamics of the decoherence factor. Moreover, we will see that this encoding is present for coupling to any site in the quasi-periodic system. To understand how the SPEs affect the decoherence factor, consider the expression for the vacuum decoherence factor $\Gamma_{i,vac}(t)$ i.e. the zero temperature limit of \eqnref{eq:dc}. Here, the dynamics in $\Gamma_{i,vac}(t)$ arises from the oscillatory term $(1-\cos(\omega_k t))$ in the sum over different SPEs with weights $\frac{|g_k^i|^2}{\omega_k^2}$. In the situation where $g_k^i$ is non-zero for many SPEs, in the long-time limit, one can expect that due to the dispersion in $\omega_k$ the oscillating terms would damp out i.e. $\Gamma_{i,vac}(t)$ would have a damped oscillatory nature and go to a constant steady state value. This can be seen explicitly for the case of the ordered tight-binding model ($\lambda = 0$ in \eqnref{eq:GAAH}), when all the SPE's are delocalized with $|g_k^i|^2=\frac{1}{N}$ and $\omega_{k,0} = \mu - 2 \cos(2\pi k/N), -N/2\leq k < N/2$ (periodic boundary condititons). In this case, we can in fact write down the steady state vacuum decoherence rate analytically as:
\begin{align}
\Gamma_{i,vac}(t\rightarrow \infty) = \frac{4}{N}\sum_{k=-\frac{N}{2}}^{\frac{N}{2}-1} \frac{1}{\omega_{k,0}^2} \stackrel{N\rightarrow \infty}{=}\frac{4\mu}{(\mu^2-4)^{3/2}} 
\end{align}
At any finite time, after an initial transient, we expect that the decoherence factor will oscillate about the steady state value with decaying oscillation amplitude. On the other hand, when $g_k^i$ is non-zero only for a few isolated values of $k$ independent of the system size $N$, we expect the dynamics of the decoherence factor $\Gamma_i(t)$ to be oscillatory and non-decaying at all times. This is precisely the case when all the SPEs are localized, for instance when $\lambda>1$ in the AAH model, since in this case $g_k^i$ is non-zero only for the SPEs localized at or near the $i^{\mathrm{th}}$ site. Before we present the results from the numerical analysis of the dynamics to back-up this intuitive picture, we would like to make some pertinent observations on how we chose the constant on-site potential $\mu$. In the rest of the paper, we chose $\mu$ such that all the eigenenergies are positive $\omega_k>0$. This guarantees that the average occupation numbers in the initial thermal state are positive. Moreover, our choice is also motivated by the fact that such a positive energy spectrum precludes the occurrence of transitions from purely Markovian to non-Markovian dynamics (by tuning $\mu$) explored in \cite{Milazzo2019}. Such transitions, explored for the simple tight-binding lattice ($\lambda = 0$) in \cite{Milazzo2019}, are caused by the nature of the energy spectrum \emph{i.e.} whenever the spectrum is continuous and has a zero energy, the dynamics in the long-time limit is entirely dominated by the zero mode leading to monotonically increasing $\Gamma_{i,\mathrm{vac}}(t)$ and Markovian dynamics, while this behaviour is absent when the spectrum does not contain zero energy modes. In contrast, we would like to examine the change in the degree of non-Markovianity of the qubit dynamics as the nature of the SPEs is changed.

We have confirmed the above mentioned behavior of the decoherence factor for the AAH model as shown \figref{fig:figure3} (a) where we show that the dynamics of the vacuum decoherence factor $\Gamma_{i,vac}(t)$ of a qubit coupled to the $i^{\mathrm{th}}$ site of the chain goes from fast damped oscillatory behaviour when $\lambda < 1$ (all SPEs delocalized) to undamped oscillatory behavior when $\lambda > 1$ (all SPEs localized). In this manner the decoherence dynamics provides a clear signature of the behaviour of the SPEs. Moreover, in order to confirm that decoherence dynamics is governed by number of SPEs that are strongly participating in the dynamics we define a quantity $P_i \equiv \sum_{k=1}^{N} |g_{k}^i|^4$ for each site  of the GAAH chain inspired by the usual inverse participation ratio used in studies of localization \cite{GAAH_mobility_edge,ipr1} The scaling of this quantity with the system size should capture the number and strength of of eigenmodes participating in the coupling to the qubit at a given site $i$. For the delocalized case, since $|g_{k}^i|^2\approx 1/N$ for all the modes we expect that $P_i \sim 1/N$. On the other hand, for the localized case, since only one mode will be strongly coupled to the qubit, then $P_i \sim N^0$. We have confirmed this behaviour in \figref{fig:figure3} (b) and in addition we see there that for the critical case with $\lambda  = 1$ in the AAH model, as expected we see intermediate behaviour with $P_i \sim  N^{-b}$ with $(0<b<1)$.
\begin{figure}
	\includegraphics[scale=0.85]{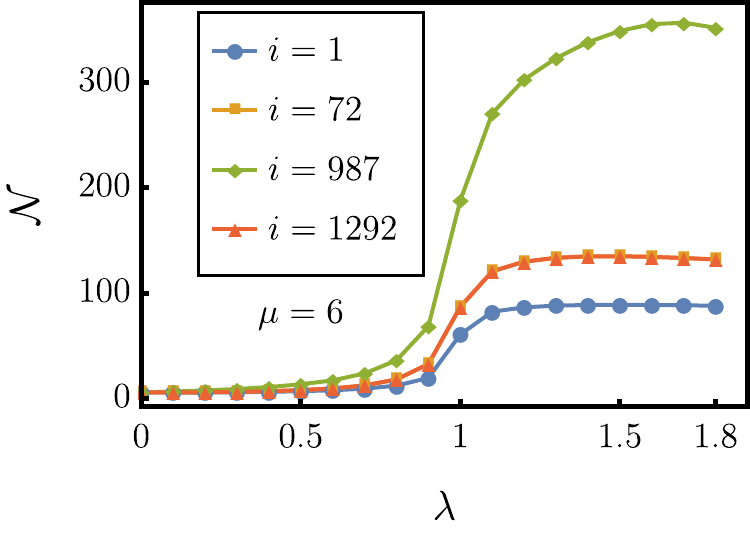}
	\caption{(Color online). Backflow of information $\mathcal{N}$ measuring the non-markovianity in the decoherence dynamics of a qubit coupled to different sites (see legend) of a quasi-periodic lattice (regular AAH model $\alpha = 0$) as a function of $\lambda$ (number of sites $N=2584$). Clearly the non-markovianity measure is sensitive to the localization-delocalization transition at $\lambda=1$.} 
	\label{fig:figure4}
\end{figure}

\begin{figure}
	\includegraphics[width=\linewidth]{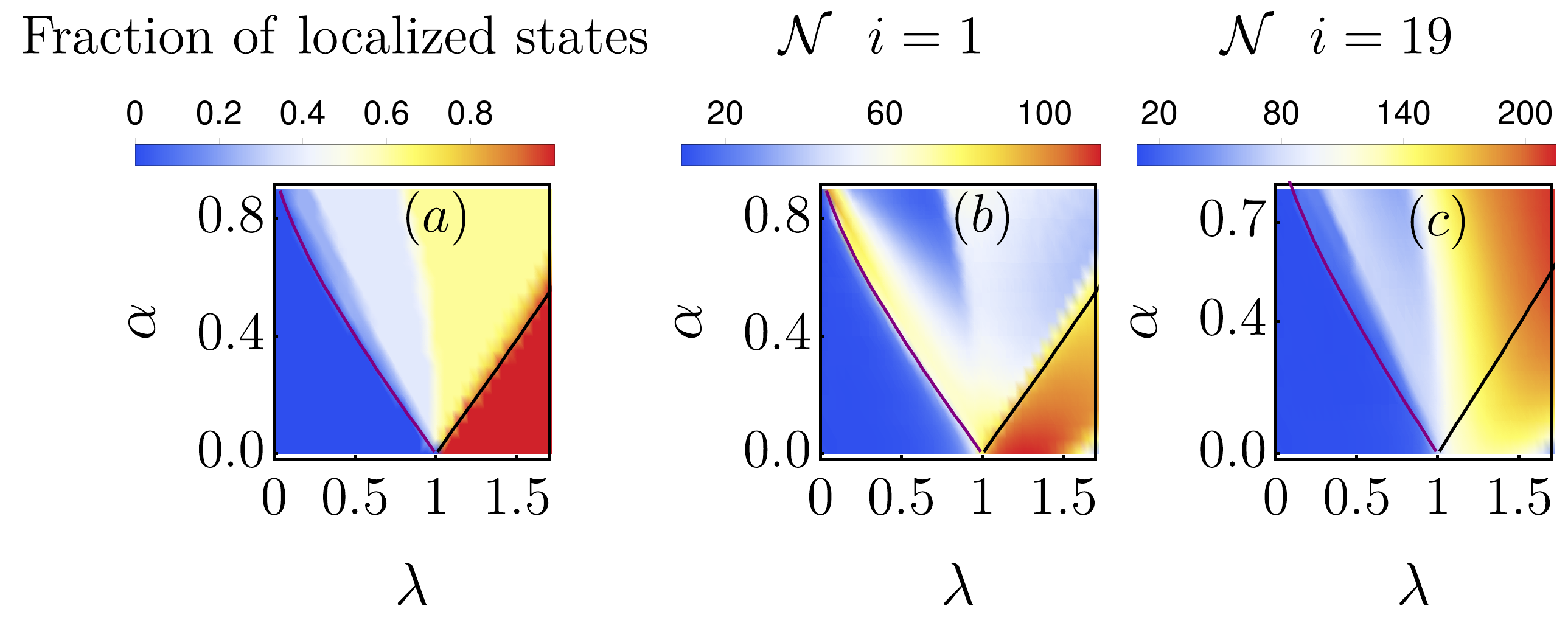}
	\caption{(Color online). (a). Fraction of localized states as a function of $\alpha-\lambda$ for GAAH model. The critical lines in purple and black separate the region with a mobility edge from the delocalized and localized regimes respectively. Backflow of information $\mathcal{N}$ as a function of $\alpha-\lambda$ for a GAAH lattice with number of sites $N=2584$ with a qubit coupled to site $i=1$ (b) and $i=19$ (c). When there is a mobility edge $\mathcal{N}$ becomes dependent on the site to which the qubit is coupled.}
	\label{fig:figure6}
\end{figure}

From an open quantum systems point of view the quasiperiodic system is essentially a bath for the qubit and the undamped oscillatory behaviour of the decoherence function is an indicator of highly non-Markovian dynamics \cite{Breuer2009,Rivas2010,Wimann2012,Lorenzo2013,Khurana2019,Milazzo2019}. Hence, the backflow of information based measure for non-markovianity developed in \cite{Breuer2009,Rivas2010,Wimann2012,Lorenzo2013} should clearly capture the difference between the dynamics in the localized and delocalized case. The essential idea of this measure is that the rate of change of decoherence factor with time, $\gamma (t)=\dot{\Gamma}(t)/2$ \citep{Khurana2019,Milazzo2019} will be negative whenever there is backflow of information from the bosonic system to the qubit. From the behaviour of $\Gamma_{vac}(t)$ in \figref{fig:figure3} (a), we can deduce that $\gamma_{vac}(t)$  will take larger negative values in a sustained manner for the localized regime ($\lambda>1$) in comparison with the delocalized regime ($\lambda<1$) showing that the backflow of information is larger in the former case. A quantitative measure of the backflow of information is given by:
\begin{align}
\mathcal{N} = \sum_{p=1}^{N_{max}} (e^{-\Gamma(t_p^f)}-e^{-\Gamma(t_p^i)}) \label{eq:BFI},
\end{align}
where the sum is over all intervals $\left[t_p^i,t_p^f \right]$ over which $\gamma(t)<0$. This measure is plotted in \figref{fig:figure4} for the AAH model at finite temperature $(\beta=1)$ and we can clearly see a sharp delocalization-localization transition from backflow of information irrespective of which site the qubit is coupled. We have also checked that this feature will be present at any finite temperature. 
Furthermore, we note that in order to calculate $\mathcal{N}$ at different values of $\lambda$ we have tracked the dynamics of $\Gamma(t)$ upto $t=1200$ with time discretisation of $\Delta t = 0.1$ for a system size of $N=2584$. 
In this manner the measure $\mathcal{N}$ can be used to clearly evidence the localization-delocalization transition.

So far we have discussed the AAH model where all the states are either delocalized, localized or critical. We now consider the GAAH model and explore the qubit dynamics when the quasi-periodic system has a mobility edge. In the presence of a mobility edge, there is a coexistence of localized and delocalized states in the spectrum. In \figref{fig:figure6} (a), we have shown the fraction of localized states in $\alpha-\lambda$ parameter space with $\alpha, \lambda>0$. In this phase diagram, there are two lines where mobility edge exactly matches with the lowest and highest eigenvalue of the system respectively. These lines are known as the ``critical lines". Below and above the two ``critical lines", all the states are either delocalized or localized respectively which means there is no mobility edge. In the rest of the regimes, there is a mobility edge with a finite fraction of localized states. This leads us to the explore the following question: how much of the `phase diagram' of the SPEs in the  GAAH model depicted in \figref{fig:figure6}(a) is captured by the backflow of information parameter $\mathcal{N}$? To answer this, we have plotted $\mathcal{N}$
in \figref{fig:figure6} (b) and (c) in the $\alpha-\lambda$ plane when the qubit is coupled to the sites $1$ and $19$ respectively. We can clearly see that in the presence of all delocalized states backflow of information is less and in presence of all localized states backflow of information is large irrespective of which site the qubit is attached. On the other hand, in the regime where there is mobility edge, the backflow of information quantified by $\mathcal{N}$ shows significant dependence on the site where the qubit is attached. While we have depicted the exemplary case for two sites, we have checked this site dependence extensively. This dependence can be justified by appealing to the fact that in this regime, the dynamics of a qubit coupled to a given site is going to depend rather strongly on whether any of the SPEs above the mobility edge are localized at or near the site in question. 
While the calculations of $\mathcal{N}$ in \figref{fig:figure6} were performed for $\beta = 1$, we have checked that the qualitative features described here are rather insensitive to temperature.
\begin{figure}
	\includegraphics[width=\linewidth]{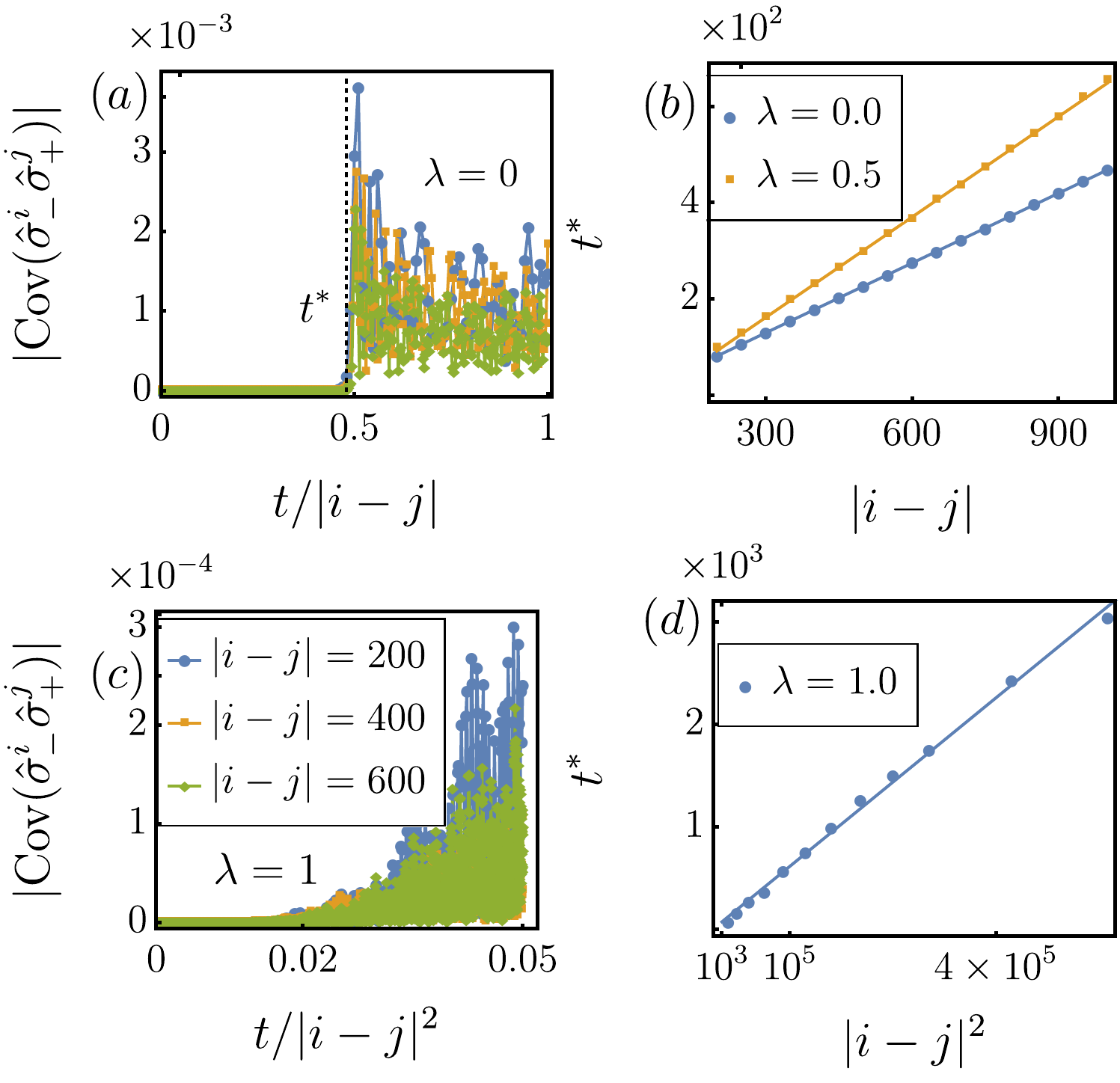}
	\caption{(Color online). Dynamics of correlation between two qubits coupled to sites $i$ and $j$ of the regular AAH model, as measured by the magnitude of the covariance, is shown for the delocalized regime ($\lambda<1$) in (a) and critical regime in (b) ($\lambda=1$). In both cases the correlations start building in a significant manner after a threshold time $t^*$. While $t^*$ scales linearly with the separation $\vert i- j \vert$ in the delocalized regime as shown in (c), in the critical regime it displays quadratic scaling - see (d). Here, we have fixed $i=N/4$, $j=3N/4$ varied the system size as $N=400, 800$ and $1200$ respectively to change $\vert i - j \vert$.}
	\label{fig:figure7}
\end{figure}

\begin{figure}
	\includegraphics[width=\linewidth]{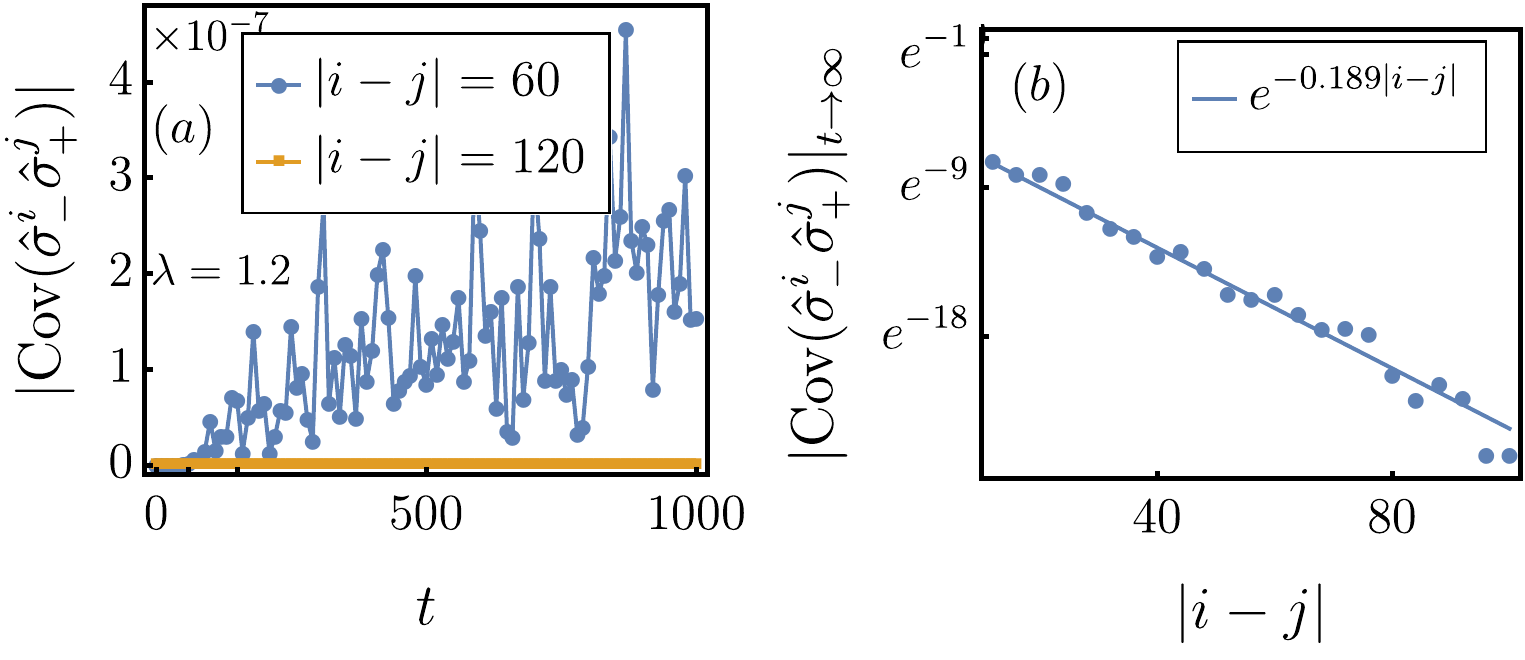}
	\caption{(Color online). Dynamics of correlation between two qubits coupled to sites $i$ and $j$ of the open regular AAH model with system size $N=400$, as measured by the magnitude of the covariance, for the localized regime with $\lambda=1.2$. (a) Exponentially decaying magnitude of correlations build up when the two sites are comparable the localization length (blue curve). Here, we have taken $i=170$ and $j=230$. When the separation is much larger than the localization length no correlations build even after long time (yellow curve). Here, we have chosen $i=140$ and $j=260$. (b) Here, we have shown the long-time ($\sim 10^4$) correlation decays exponentially with the distance between two qubits in the localized regime.}
	
	\label{fig:figure7a}
\end{figure}

\begin{figure}
	\includegraphics[scale=0.85]{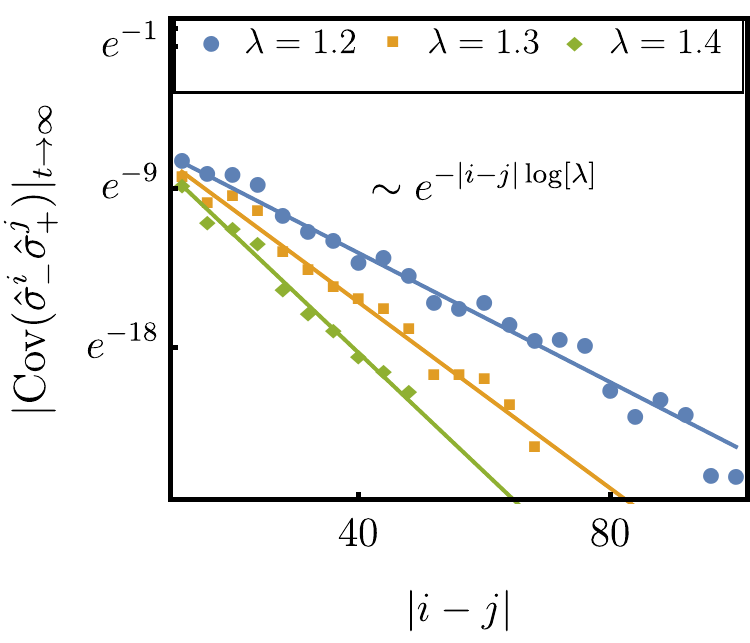}
	\caption{(Color online). Magnitude of covariance at long times ($\sim 10^4$) is plotted as a function the distance $i-j$ between two qubits in the localized regime of regular AAH model with $\lambda=1.2$ (blue dots), $~1.3$ (orange squares) and $1.4$ (green squares). Clearly the exponential decay of the covariance agrees with the straight lines which represent an exponential decay of the form $\exp(- \vert i - j \vert \log[\lambda])$ expected from the behavior of localized SPEs.}
	
	\label{fig:figure7b}
\end{figure}

\begin{figure*}
	\includegraphics[width=\linewidth]{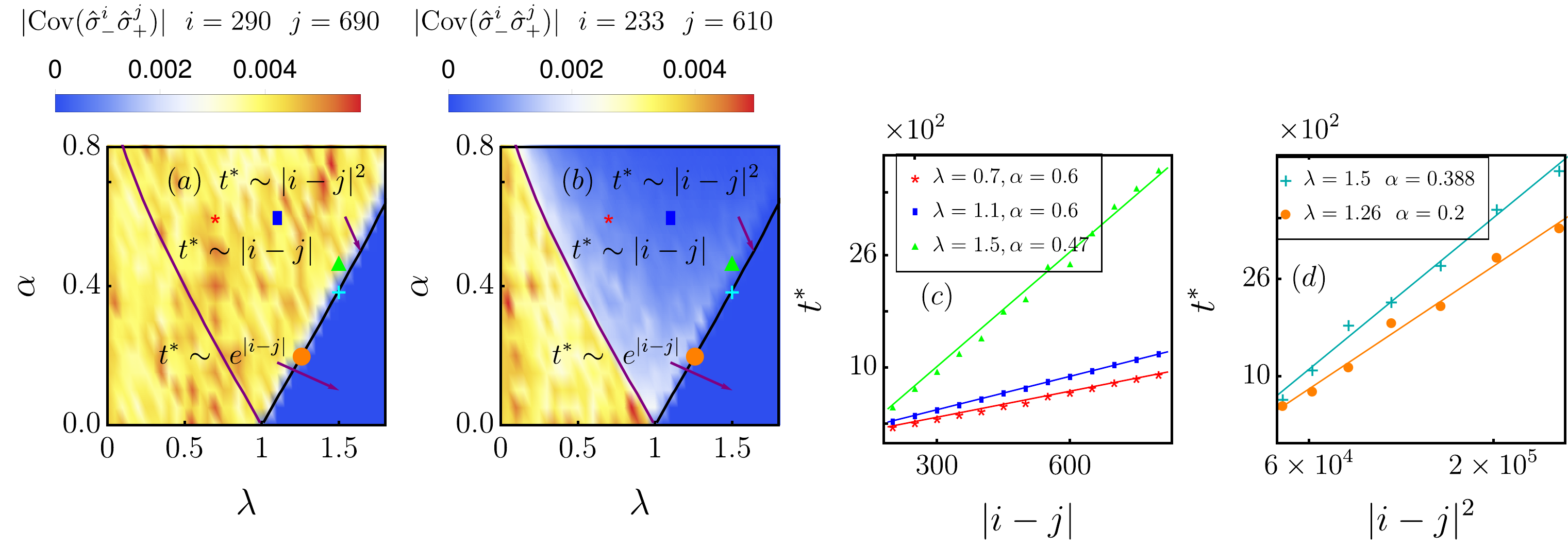}
	\caption{(Color online). Correlation strength in the long-time limit for two qubits coupled to sites $i$ and $j$ of the GAAH model in the $\alpha-\lambda$ plane. In (a), [$i = 290,j=690$], and in (b) [$i = 233, j = 610$] the correlation strength is large (small) in the fully delocalized (localized) regimes with no mobility edge. In contrast, the correlation shows dependence on the site to which the qubits couple when there is a mobility edge. In (a) and (b) system size $N=2584$. In (c) [(d)] we show the linear [quadratic] scaling of the threshold time $t^*$ as a function of qubit separation for parameters above the black critical line [on the black critical line]. Here, we have fixed $i=N/4$, $j=3N/4$ varied the system size as $N$ to change $\vert i - j \vert$.}
	\label{fig:figure8}
\end{figure*}
\subsection{Two qubits coupled to a quasi-periodic system}
We now consider the dynamics of the covariance factor $\mathrm{Cov}(\hs_-^i \hs_{+}^j)$ (see \eqnref{eq:CovExpression}) as a measure of the correlations between two qubits coupled to sites $i$ and $j$ of the quasi-periodic system. We begin with the AAH model ($\alpha = 0$ in \eqnref{eq:GAAH}) and plot the dynamics in the three distinct regions $\lambda < 1$ (delocalized), $\lambda = 1$ (critical), and $\lambda>1$ (localized) in \figref{fig:figure7} (a), (c), and \figref{fig:figure7a} respectively. In Fig. \figref{fig:figure7} (a) and (c) we can clearly see that after a threshold time $t^{*}$ correlations develop between the two qubits as evidenced by the non-zero values of the covariance. Moreover, for $\lambda<1$ when all the SPEs are delocalized the threshold time scales as $t^*\sim |i-j|$ and when $\lambda=1$ with SPEs critical, $t^*\sim |i-j|^2$ as shown in \figref{fig:figure7} (b) and (d) respectively. For $\lambda>1$, all the SPEs are exponentially localized and the localization length is given by $\xi=1/\log[\lambda]$ \cite{Modugno2010} (recall that $\lambda$ is in units of $J$ in our paper). In this case, when the distance between two qubits is comparable to the localization length, after some threshold time $t^*$, a small but discernible amount of correlation builds up as shown in \figref{fig:figure7a} (a) (blue curve). But as shown in \figref{fig:figure7a} (b) the magnitude of this correlation, even at long times ($\sim 10^4$), is exponentially suppressed as a function of the distance between the qubits and as a result for distances much larger than the localization length, correlations remain close to zero for extremely long times (see yellow curve \figref{fig:figure7a} (a)). In order to clearly demonstrate that the localization length implied by the exponential decay of the magnitude of the long-time correlations with separation between qubits is precisely the same as that expected from the behavior of the SPEs given by $1/\xi=\log[\lambda]$, we have plotted the covariance magnitude at long-times as a function of distance between two qubits for three different values of $\lambda$ in \figref{fig:figure7b}. In this manner we can explicitly read-out the localization length from the correlation dynamics of the attached qubits. Thus, as we anticipated earlier the behavior of the correlation dynamics is completely analogous to the diffusion dynamics of an initially localized wavepacket with the role of the wave-packet width played by the distance between the qubits and the time of spread represented by the threshold correlation time $t^*$. In this manner the correlation dynamics provides a very clear signature of the transport properties in the AAH chain.

Let us now consider the dynamics of the covariance in the presence of a mobility edge, which is a possibility when the qubits are coupled to the GAAH chain. In \figref{fig:figure8} (a) and (b) we have plotted the long-time value of the covariance of the qubits coupled to two different pairs of sites in the $\alpha-\lambda$ plane of GAAH model. In both \figref{fig:figure8} (a) and (b), the region with all SPEs localized below the critical line (with $\lambda>1$) is clearly distinguished with zero covariance as we have chosen the distance between the sites much larger than the localization length. In contrast, in the region above the critical lines of the phase plots in \figref{fig:figure8} (a,b) where there is a finite fraction of delocalized states with a mobility edge in the spectrum, the magnitude of the covariance depends on the coupling sites of the qubit 
Indeed the strong dependence on the site to which the qubits are coupled is somewhat similar to the behaviour of the backflow of information parameter $\mathcal{N}$ in \figref{fig:figure6} 
Finally, we see that below the left critical line in the region without mobility edge the magnitude of the covariance is significant and independent of the coupling sites. Coming to the scaling behaviour of the threshold time $t^*$ as a function of the distance between the sites $\vert i - j \vert$, we find linear scaling whenever there is a finite fraction of delocalized states as shown in \figref{fig:figure8} (c). Interestingly, we also see there that as we move into regions with larger fraction of localized states (left to right in the $\alpha-\lambda$ plane) the slope of the linear scaling $t^* \sim \vert i-j\vert$ becomes larger indicating that it takes longer for the correlations to set in. In \figref{fig:figure8} (d) we consider the scaling for two points on the critical line with $\lambda > 1$ where there is a mixture of critical and localized states and clearly see that $t^* \sim |i-j|^2$ i.e. the scaling is governed by the nature of the states that help set up correlations at the fastest rate. Moreover, we find that the scaling behaviour we observe is rather insensitive to temperature. This can be anticipated from the fact that the delocalised and critical states always have lower energy than the localized states. Thus, using a combination of the magnitude of the correlation and the scaling behaviour of the threshold time for correlation of the two qubits, we can clearly extract the transport properties of the quasiperiodic GAAH chain. 
	

\section{Conclusions}

In conclusion, we have presented a theoretical scheme to read-out the nature of on-site potential, single particle states and isolated system transport properties of a non-interacting quasi-periodic system by coupling it to probe qubit systems. A single qubit coupled to any site of the system shows strikingly different decoherence dynamics depending on the presence of all delocalized or all localized states in the quasi-periodic system. This difference in dynamics is quantifiable via the backflow of information measure for the non-markovianity and captures delocalization-localization transition of the regular AAH model upon changing $\lambda$ (with higher backflow of information in the localized regime). In the GAAH model, in the presence of a mobility edge, we find that the backflow of information is site-dependent for a given $\lambda$ and $\alpha$. Depending on the number of localized and delocalized states, the dynamics will show high and low backflow at different sites. Nonetheless, we see that there are multiple sites in the lattice upon coupling to which with a one qubit probe we can read out the the phase diagram of the GAAH model in-terms of the fraction of localized states in the $\alpha-\lambda$ parameter plane. When the two qubits are coupled at two distinct sites $i$ and $j$, we were able to show that there is a threshold time $t^*$ after which correlations develop between initially uncorrelated qubits. More interestingly, the scaling of this threshold time as a function of the distance between the qubits $\vert i - j \vert$ contains the signature of the transport properties expected in the quasi-periodic system. We have shown that in the regular AAH model when all the states are delocalized, corresponding to ballistic transport, we obtain $t^*\sim |i-j|$, with all the states are critical with  diffusive transport we get $t^*\sim |i-j|^2$, and in the localized regime with no transport we see the scaling $t^*\sim exp(|i-j|)$. In the presence of a mobility edge in the GAAH model with the coexistence of SPEs of different nature, the scaling is dominated by the fastest states. For instance, with a mixture of delocalized and localized states, scaling is governed by the delocalized states. In this manner we are able to again extract a phase plot of the GAAH model in terms of the different transport behaviour expected. In general we found that the initial temperature of the quasi-periodic system does not qualitatively affect our results.

Finally, let us examine the prospects for experimentally realising the theoretical scheme we have proposed. Focusing first on the decoherence dynamics of one qubit coupled to a quasi-periodic chain, we note that multiple elements required to implement this are already in place especially in ultracold atomic systems. This includes realizations of quasi-periodic AAH and GAAH lattices \cite{expt1,GAAH_experiment,Nagler2020} and experiments with position controlled implantation of qubit impurities in ultracold gases and studies of their decoherence \cite{Schmidt2018,Ratschbacher2013}. In \cite{Schmidt2018} although the coupling between the impurity qubit and the atomic gas was implemented via elastic collisions is expected to cause dephasing, for their choice of the internal atomic states for the qubit, dephasing was dominated by other external noise factors. Nevertheless, they do comment that this can be mitigated by choosing different sets of internal states for the qubit. Thus, in summary we anticipate that by adding a quasi-periodic optical lattice potential for the BEC in \cite{Schmidt2018}, the first part of our proposal could be readily tested. In contrast we believe that the second part of our proposal is somewhat more challenging to implement. While the position controlled implantation of two qubit impurities in an ultracold gas should be feasible following \cite{Schmidt2018}, an experimental challenge would be to implement, \emph{within the same set up}, controlled gates required to measure the correlations between the qubits to read out transport properties of the chain. We hasten to add that in general two qubit gates are routinely implemented in quantum technology platforms such as trapped ions \cite{Barreiro2011,Haffner2008,Negnevitsky2018}. One way to address this challenge would be to consider alternative platforms such as photonic waveguide arrays \cite{zilberberg1} or cavity polaritonic devices \cite{Goblot2020} where quasi-periodic lattices have been implemented and consider interfacing them with  qubits \cite{Quinteiro2006,Puri2014}. Since the degree of control needed to establish the photonic or polaritonic lattice system is less than that for ultracold atoms in optical lattices, it may be less challenging to establish the control gates to measure qubit correlations. Though, the suggestions we have made above will realize an in-situ measurement, since the qubits are immersed or interfaced into the chain, it may be also prudent to consider schemes for near-field coupling of qubits to lattices for instance via electric or magnetic dipole interactions to implement a continuous read-out. As part of future research, we will develop a concrete experimental proposal along the above mentioned lines to realise the proposed theoretical scheme.

Apart from developing a detailed proposal for the experimental proposal, there are multiple interesting theoretical questions arising from this work that would be of interest in a follow-up study. While we have focused on the impact of the quasi-periodic system on the qubit, we have not really paid attention to the back-action of the qubit on the system. This could be especially interesting in situations where the coupling is large. In the same manner, for the two qubit case we have begun with a product state of the qubits but we expect that it will also be fruitful to understand what happens to qubits that have initial correlations and whether one can learn something more from the dynamics of the entanglement between the qubits. Finally, while we have focused on the exactly solvable pure dephasing coupling between the qubits and the quasi-periodic system, it would interesting to explore if we can obtain direct signatures of transport such as the current by considering couplings that allow energy exchange between the qubit and the system \cite{Lorenzo2017}.

\section{Acknowledgement}
M.~S would like to acknowledge Archak Purkayastha for useful discussions. M.~S acknowledges funding from post-doctoral fellowship of IIT Gandhinagar with project code MIS/IITGN/PD-SCH/201415/006. B.~P.~V. is supported by the Research Initiation Grant, Excellence-in-Research Fellowship of IIT
Gandhinagar, and a Department of Science \& Technology Science and Engineering Research Board (India) Start-up Research Grant No. SRG/2019/001585. 
\appendix*
\section{Dephasing Spin-Boson Model Exact Solution} \label{app:dephSBmodel}
In this appendix, for the sake of completeness and clarity, we adapt to the setting in this paper and present the derivation of the spin-boson dynamics for dephasing baths presented in previous works \cite{MassimoPalma1996,Lidar2001,Reina2002,Breuer2007,Cipolla2018,Chen2018,Milazzo2019}. We will work out the calculation for the situation with two qubits coupled to a Bosonic bath and obtain the one qubit situation as a simple limiting case. Let us begin by rewriting the two qubit hamiltonian \eqnref{eq:spinboson_2q} as \cite{Cipolla2018}:
\begin{align}
\Hop_{2q}^{SB} = \Hop_{S}[\hsb] + \Vop[\hsb],
\end{align}
with 
\begin{align}
\Hop_S[\hsb] = \frac{\omega_A}{2} \hs_{z}^i + \frac{\omega_B}{2} \hs_{z}^j, 
\end{align}
is the free spin hamiltonian and 
\begin{align}
\Vop[\hsb] = \sum_{k=1}^{N} \omega_k \heta_k^{\dagger} \heta_k  + \sum_{k=1}^{N}  (f_k[\hsb] \heta_k^{\dagger}+ f_k^{*}[\hsb] \heta_k),
\end{align}
is the sum of free bosonic and interaction terms. In this appendix, we will use the notation $\hsb = (\hs_z^i,\hs_z^j)$ and $\esb = (\sigma_z^i,\sigma_z^j)$ for the for the $z-$ component of spin operators and eigenvalues of the qubit respectively. Note that $\sigma_z^{\tau} = \pm 1$ ($\tau = i,j$). The coupling function $f_k$ with the $k^{\mathrm{th}}$ bosonic mode is given by:
\begin{equation}
f_k[\hsb] = g_k^{i*} \hs_z^i + g_k^{j*} \hs_z^j. \label{eq:coupfunction}
\end{equation}
Our aim is to calculate the time evolution in the Heisenberg picture for the operators $\hs_{\pm}^{i}(t),\hs_{\pm}^{j}(t)$, since we already know that $\hsb$ do not evolve with time. The time evolution of an arbitrary spin operator $\Aop(t)$ is given by:
\begin{align*}
\Aop(t) &= e^{i\Hop t} \Aop(0)  e^{-i\Hop t}=e^{i\Vop[\hsb]t} \Aop_s(t) e^{-i\Vop[\hsb]t},
\end{align*}
where $\Aop_s(t) = \left ( e^{i\Hop_S[\hsb]t} \Aop (0) e^{-i\Hop_S[\hsb]}\right)$ is just the free evolution of the spin operator $\Aop$. We can simplify the problem by working in the product basis of $\hs_z^{i}$ and $\hs_z^j$, denoted by $\ket{\esb} = \ket{\sigma_z^i}_i\otimes\ket{\sigma_z^j}_j$, as:
\begin{align}
\Aop(t) &= \sum_{\esb,\esb^{\prime}} \ket{\esb^{\prime}}\bra{\esb} e^{iW_t(\esb,\esb^{\prime})} \prod_{k=1}^N \Dop_k \left(\mu_k[\esb,\esb^{\prime}]\right) \sandwich{\esb^{\prime}}{\Aop_s(t)}{\esb} \label{eq:Aopfnt}
\end{align}
with
\begin{align*}
W_t(\esb,\esb^{\prime}) &= \sum_k \left(\left\vert f_k [\esb^\prime]\right\vert^2-\left\vert f_k [\esb] \right\vert^2\right) \frac{\sin(\omega_kt)-\omega_kt}{\omega_k^2} \nonumber \\
&+2 \operatorname{Im}\left(f_k^*[\esb^\prime]f_k[\esb]\right) \frac{1-\cos(\omega_k t)}{\omega_k^2} \nonumber\\
\mu_k[\esb,\esb^{\prime}] &= \frac{f_k[\esb^\prime]-f_k[\esb]}{\omega_k}(e^{i\omega_k t}-1) \nonumber.
\end{align*} 
With $\esb = (\sigma_z^i,\sigma_z^j)$ and $\esb^\prime = (\sigma_z^{i\prime},\sigma_z^{j\prime})$ and using \eqnref{eq:coupfunction} for $f_k$, the coefficients in the above definition simplify further as:
\begin{align}
&W_t(\esb,\esb^\prime) = (\sz^{i\prime}\sz^{j\prime}-\sz^i\sz^j)\sum_k 2 \operatorname{Re}(g_k^i g_k^{j*})  \frac{\sin(\omega_kt)-\omega_kt}{\omega_k^2} \nonumber\\
& + (\sz^{i}\sz^{j\prime}-\sz^{i \prime}\sz^j) \sum_k 2 \operatorname{Im}(g_k^ig_k^{j*}) \frac{1-\cos(\omega_kt)}{\omega_k^2} \label{eq:Wt}\\
&\mu_k[\esb,\esb^{\prime}] = \left[g_k^i(\sz^{i\prime}-\sz^{i}) + g_k^j(\sz^{j\prime}-\sz^{j}) \right]\frac{e^{i\omega_k t}-1}{\omega_k} \label{eq:mukt}
\end{align}
Noting that $\left[\hs_{-,s}^{i}(t)\right]^{\dagger} = \hs_{+,s}^{i}(t) = \hs_{+,s}^{i}(0) e^{i\omega_A t} = \left( \ket{1}_i\bra{-1} e^{i\omega_A t} \right) \otimes \hat{I}_j$ and $\left[\hs_{-,s}^{j}(t)\right]^{\dagger} = \hs_{+,s}^{j}(t) = \hs_{+,s}^{j}(0) e^{i\omega_B t} = \hat{I}_k \otimes \ket{1}_j\bra{-1} e^{i\omega_B t}$, we can use \eqnref{eq:Aopfnt}, \eqnref{eq:Wt}, and \eqnref{eq:mukt} to obtain Eqs. \eqref{eq:ladderifnt} and \eqref{eq:ladderjfnt}. Setting $g_k^j \equiv 0$ in \eqnref{eq:ladderifnt} immediately gives \eqnref{eq:singlequbitSigma}. Finally, some standard results from quantum optics required to derive Eqs. \eqref{eq:CovExpression}, \eqref{eq:dc}, and \eqref{eq:corrdc} are $\Dop_k(\alpha) \Dop_k(\alpha^\prime) = \Dop_k(\alpha+\alpha^\prime)e^{-(\alpha \alpha^{\prime*}-\alpha^{\prime} \alpha^{*})/2}$ and the thermal state average of displacement operators:
\begin{align*}
\frac{\Tr \left[ \prod_{k=1}^N \Dop_k (\alpha_k) e^{-\beta \omega_k \hetad_k(0) \heta_k(0)} \right]} {Z_{\beta}} = e^{-\sum_k \vert \alpha_k \vert^2 \coth(\beta \omega_k/2)}.
\end{align*}
%

\end{document}